\documentclass[aps,prb,twocolumn, groupedaddress,showpacs]{revtex4}
\usepackage{graphicx,amsmath}
\usepackage{amssymb}

\begin{document}
\title{Study of the Hubbard model on the triangular lattice using dynamical
  cluster approximation and dual fermion methods}

\author{Hunpyo Lee}
\author{Gang Li}
\author{Hartmut Monien}
\affiliation{Physikalisches Insititut, Universit\"at Bonn, 53115 Bonn, Germany}
\date{\today}

\begin{abstract}
  We investigate the Hubbard model on the triangular lattice at half-filling
  using the dynamical cluster approximation (DCA) and dual fermion (DF) methods
  in combination with continuous-time quantum Monte carlo (CT QMC) and
  semiclassical approximation (SCA) methods. We study the one-particle properties and
  nearest-neighbor spin correlations using the DCA method. We calculate
  the spectral functions using the CT QMC and SCA methods. The spectral
  function in the SCA and obtained by analytic continuation of the Pade
  approximation in CT QMC are in good agreement. We determine the
  metal-insulator transition (MIT) and the hysteresis associated with a
  first-order transition in the double occupancy and nearest-neighbor spin
  correlation functions as a function of temperature. As a further check, we
  employ the DF method and discuss the advantages and limitation of the
  dynamical mean field theory (DMFT), DCA and recently developed DF methods by
  comparing Green's functions. We find an enhancement of antiferromagnetic (AF)
  correlations and provide evidence for magnetically ordered phases by
  calculating the spin susceptibility.   
\end{abstract}
 
\maketitle

% Main body
\section{Introduction\label{Introduction}}

The physics of systems which exhibit strong electronic correlations and geometric
frustration at the same time is still unclear and therefore interesting. 
Recent experiments, such as discovery of the pyrochlore compound ${\rm
  LiV_{2}O_{4}}$ \cite{Kondo97} which show heavy fermion behavior
and organic materials $\kappa$-(BEDT-TTF)$_{2}$X \cite{RH97} which exhibit various interesting phases,
motivated us to study the frustrated systems in more detail. Theoretically they
are described by a two-dimensional one-band Hubbard or t-J models
on the triangular lattice. It is well known that on the square lattice at
half-filling the ground state is a Mott insulator with AF order
but on the triangular lattice the frustration suppresses AF order and we
expect to find a Mott transition. 

In this paper, we study the model which was presented in a recent paper of Imai and Kawakami \cite{Imai02}. 
They used the DCA method \cite{Hettler98, Hettler00} in combination with
noncrossing approximation (NCA) and fluctuation exchange (FLEX) methods
at high temperature regions in metallic states to demonstrate
how geometrical frustration suppresses AF correlations by
tuning aniotropic hopping t' in Fig. 1(a). However, the methods used in the
paper are limited to high temperatures. For this reason we investigate
the low temperature MIT with a first-order transition and the evidence
of magnetically ordered phases. In addition, we test the newly developed DF method
\cite{Rubtsov06, Brener07, Gang08}
beyond the single-site DMFT method by comparing the Green's function of
single-site DMFT \cite{Georges96,Metzner89}, DF and DCA methods with $N_c$=4 and $N_c$=16. 
 
The paper is organized as follows: In Sec. \ref{Formalism}, we introduce the
model and discuss the advantages and limitations of the computational methods briefly. In
Sec. \ref{Results}, we present the results. In the first part, we compare the spectral functions which
exhibit the quasiparticle peak and gap structure obtained by the SCA \cite{Okamoto05,Fuhrmann07} and
CT QMC \cite{Rubtsov05} methods with Pade approximation. In the
second part, we show that the MIT is a first-order
transition by measuring the total density of state (DOS), double occupancy and
nearest-neighbor spin correlations. In the third part, we calculate the
single-particle Green's function using the DMFT, DF and DCA methods with $N_c$=4 and
$N_c$=16. Specifically, we show that the DF method which is based on the
single-site DMFT method can describe non-local correlation effects very well. In
the last part, we explore the spin susceptibility using the DF
method. Sec \ref{Summarize}, we give a summary work.  
 
%% second section
\section{Model and Methods}\label{Formalism}
\subsection{Model}
 We consider the two-dimensional Hubbard model on the triangular lattice. 
\begin{equation}
  H=-t\sum_{\langle i,j \rangle \sigma}c^{\dag}_{i\sigma}c_{j\sigma} 
+ U \sum_{i}n_{i\uparrow}n_{i\downarrow}
\end{equation} 
where $c_{i\sigma}(c^{\dag}_{i\sigma})$ is the annihilation (creation) operator
of an electron with spin $\sigma$ at the $i$-th site, t is the hopping matrix
element and $U$ represents the Coulomb repulsion. In this paper we only
consider the isotropic hopping of t'=t in Fig. 1(a)-(b).

\begin{figure}[htb]
\includegraphics[width=170pt]{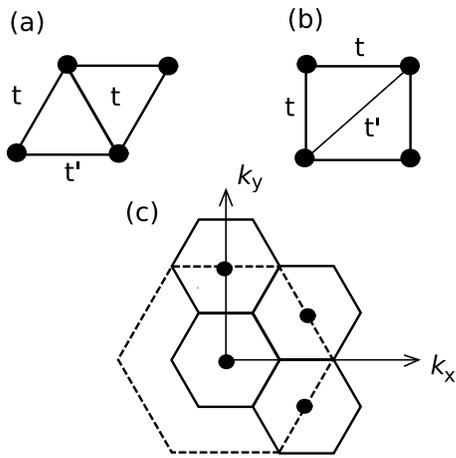}
\caption {(a) Schematic representation of triangular lattice with electron
  hoppings. (b) Equivalent representation of (a) for a square structure. (c) Example of the coarse-graining cells in the BZ for the triangular
  lattice (a), where the cluster size is $N_c$ = 4.}
\label{fig:BZ}
\end{figure} 

Due to the geometrical frustration, this model has broken particle-hole symmetry
even at half-filling unlike the case of the square lattice and the original
Brillouin zone (BZ) has an hexagonal structure shown by the doted line in
Fig. 1(c). For this model there are a lot of studies using a variety of methods
such as the path integral renormalization
group method \cite{Morita02}, the quantum Monte carlo method \cite{Bulut05},
the DMFT \cite{Aryanpour06, Merino08} and the
cluster extension method of DMFT \cite{Parcollet04, Kyung07, Ohashi08,
  Kyung06}. Especially, DF \cite{Rubtsov06, Brener07,
  Gang08} and DCA methods \cite{Hettler98, Hettler00} are noteworthy because
both methods can capture non-local correlation effect which
are lost in the single-site DMFT and they are computational cheaper and have
less of a sign problem than the lattice QMC calculation. In short,
DCA method can treat correlations up to a cluster size $N_c$ accurately and
long range correlations are considered on the mean-field level. On the other
hand, the DF method considers long range as well as short range correlations within
pertubative diagram expansion, which is done by introducing an auxiliary field. Because each
method has its limitations, it is useful to compare results of both. We use the CT QMC method
\cite{Rubtsov05}, which can access the low temperature region easily without
the Trotter error, as well as the SCA method \cite{Okamoto05, Fuhrmann07} as
impurity solvers. 

\subsection{DCA method}
The DCA method \cite{Hettler98,Hettler00} assumes that the self-energy in the
first BZ is constant and the coarse-grained Green's
function (DCA equation) is given by Eq. (2).
\begin{eqnarray}
\overline{G}_{\sigma}({\bf K},z)=\frac{1}{N}\sum_{\tilde{\bf K}} 
\frac{1}{z-\epsilon_{{\bf K+\tilde{K}}}-\Sigma_{\sigma}({\bf K},z)},
\end{eqnarray}
where N is the number of lattice sites in each first BZ and the summation over ${\bf
  \tilde{K}}$ is calculated in each of them. For delimitation we consider an
  example of $N_c$=4 in order to explain the DCA method. The
first BZ (dashed line in Fig. 1(c)) is created by partitioning the original BZ.
Like the standard procedure of DMFT, the coarse-grained Green's function is determined self-consistently after
  several iterations. The main advantages of the DCA method are that it considers short
  range correlations in the cluster size exactly and has smaller computational
  load and fermionic sign problem compared to the lattice calculation by QMC
  method. However, it is still expensive in terms of computational time and long range correlations are just
  treated on the mean-field level.   

\subsection{DF method}
 The DF method \cite{Rubtsov06,Brener07,Gang08} is a relatively new method which can describe non-local correlations based
 on the single-site DMFT method. The basic idea of the DF method is to convert
 the hopping of different fermions into an effective coupling to an
 auxiliary field. Each lattice site can be viewed as an impurity which is
 easily described by the DMFT method. While these impurities are not totally
 isolated, they are perturbatively coupled by the auxiliary field. The starting point is
 the action of DMFT which is represented in the form 

\begin{eqnarray}
S[c^{+},c]=\sum_{i}S_{imp}^{i}-\sum_{\nu,k,\sigma}(\Delta_{\nu}-\epsilon_{k})c_{\nu
  k\sigma}^{\dagger}c_{\nu k\sigma}^{\phantom\dagger}
\end{eqnarray} 
where $\Delta_{\nu}$ is the hybridization function describing the interaction
of an effective impurity with a bath and $\nu$ is the fermionic Matsubara
frequency. Here we use the dispersion relation
$\epsilon_{k}=-2t[\cos(k_{x})+\cos(k_{y})+\cos(k_{x} + k_{y})]$ based on the
correspondence of a triangular lattice to a square lattice with diagonal
hopping (b) in Fig. 1. 
This ``square lattice'' has a simple BZ which makes the
momentum summation to be easily performed by using the fast Fourier
transformation (FFT).
By the dual transfomation, the lattice problem is changed to an impurity problem which
is coupled by the auxiliary field $f(f^{\dagger})$. 

\begin{eqnarray}
S[c^{\dagger},c;f^{\dagger},f]&=&\sum_{i}S_{imp}^{i}+\sum_{k,\nu,\sigma}[g_{\nu}^{-1}(c_{k\nu\sigma}^{\dagger}f_{k\nu\sigma}^{\phantom\dagger}+h.c)
\nonumber\\
&&\hspace{1cm}+g_{\nu}^{-2}(\Delta_{\nu}-\epsilon_{k})^{-1}f_{k\nu\sigma}^{\dagger}f_{k\nu\sigma}^{\phantom\dagger}]
\end{eqnarray}
The lattice Green's function is derived from the exact relation between
Eq. (3) and Eq. (4). 

\begin{eqnarray}
G_{\nu, k}=g_{\nu}^{-2}(\Delta_{\nu}-\epsilon_{k})^{-2} G_{\nu,k}^{d}+(\Delta_{\nu}-\epsilon_{k})^{-1}
\end{eqnarray}
where $G_{\nu,k}^{d}$ is the dual Green's function and $g_{\nu}$ is the local
Green's function calculated by single-site DMFT. The main point for this
method is that in Eq. (4) the integration over $c^{\dagger}$ and c can be
performed separately for each site which yields an effective action of the
auxiliary field $f$ and $f^{\dagger}$. The Taylor expansion in powers of
$f^{\dagger}$ and $f$ will introduce the two, three, $\dots$ ,-particle vertex
functions. Using the skeleton-diagram expansion we calculate the dual
self-energy and dual Green's function by the Dyson equation. We obtain the
lattice Green's function via Eq. (5). 
Even though the DF method is an approximate method, it considers not only the short range but also the long
range correlations. Moreover, the calculation of the
two-particle properties does not introduce serious computational burden and
fermionic sign problem. 
   
\subsection{SCA method}
At high temperature the Monte carlo integration over the auxillary classical
 field $\phi(\tau)$ can be approximated by assuming $\phi(\tau)\approx$
 const. This approximation is useful because it allows to check the QMC
 results at temperature quickly. In this part we introduce the SCA method \cite{Okamoto05,Fuhrmann07} as
 impruity solver for DCA method. We consider a four-site cluster($N_c$=4) for
 triangular lattice like the structure of Fig. 1(a). In this case the
 partition function is defined as a functional integral over
 2$\times$4-component spin and site-dependent spinor fields $c^{\dagger}$ and c
as

\begin{equation}
Z=\int{\cal D}[c^{\dagger}_i c_i]e^{-S_{eff}},
\end{equation}
where
\begin{equation}
\begin{split}
S_{eff}=&\int_0^\beta d\tau\int_0^\beta d\tau'd\tau~c^\dagger(\tau){\bf
  a_{\sigma}}(\tau,\tau')c (\tau')\\
&+\int_0^\beta d\tau
  \sum_{i=0}^{N-1}Un_{i,\uparrow}(\tau)n_{i,\downarrow}(\tau),
\end{split}
\end{equation}

Here ${\bf a_{\sigma}}(\tau,\tau')$ is the Weiss field which is determined
self-consistently by Eq. (2) and $\beta$=1/T is the inverse temperature. In
this model ${\bf a_{\sigma}}$ is given as 

\begin{displaymath}
\mathbf{\bf a_{\sigma}}(\tau,\tau') = 
\left( \begin{array}{cccc}
a_{0\sigma} & a_{1\sigma} & a_{1\sigma} & a_{1\sigma} \\
a_{1\sigma} & a_{0\sigma} & a_{1\sigma} & a_{1\sigma} \\
a_{1\sigma} & a_{1\sigma} & a_{0\sigma} & a_{1\sigma} \\
a_{1\sigma} & a_{1\sigma} & a_{1\sigma} & a_{0\sigma}
\end{array} \right)
\end{displaymath}
We can decouple the interaction term as
\begin{equation}
Un_{i\uparrow}(\tau)n_{i\downarrow}(\tau)=\frac{U}{4}\left[N_i^2(\tau)-M_j^2(\tau)\right],
\end{equation}
with $n_\uparrow n_\downarrow=\frac{1}{4}\left((n_\uparrow+n_\downarrow)^2-(n_\uparrow-n_\downarrow)^2\right)=\frac{1}{4}(N^2-M^2)$.
We employ the continuous Hubbard-Stratonovich(HS) transformation in order to decouple M
terms related to auxiliary field $\phi_j(\tau)$. Here we assume that
$\phi_j(\tau)$ is $\tau$ independent and N term is neglected because charge
fluctuations are small at half-filling. By a Grassmann integration we can rewrite the partition function which is
represented as a four-dimensional integration in terms of $\phi_j$ and the fermionic
Matsubara frequency.

\begin{equation}
Z=\int d{\vec{\phi_{j} }}e^{-S_{eff}[{\bf a}(i\omega),\phi_{j}]},
\end{equation}
where the effective action $S_{eff}=\beta V$ is defined by
\begin{equation}
V(\vec{\phi})=\frac{{\phi_{1}}^2 + {\phi_{2}}^2 + {\phi_{3}}^2 + {\phi_{4}}^2}{U}-T\sum_{\omega_{n},\sigma}\ln
  \det[-\beta {\cal M}]
\end{equation}
where ${\cal M}$ is defined as
\begin{equation}
{\cal M} = {\bf a} + \hat{1}\phi_{j}\sigma_{z}
\end{equation}   
Here j=1,2,3,4 and $\sigma_z$ is the z-component Pauli matrix. The impurity Green's function is calculated by
\begin{equation}
G_j = \frac{1}{N_c} \frac{\delta \ln Z}{\delta a_j}
\end{equation}
In real frequency space the spectral function is calculated by replacing ${\bf
  a}(i\omega)$ to ${\bf a}(\omega)$. The SCA method is not only cheap in  computational time but also gives good
results in the strong coupling regime. On the other hand, it underestimates the spectral
function around $\omega=0$ and gives qualitatively wrong results at low temperature.  

\subsection{CT QMC method}
Here we describe the CT QMC method \cite{Rubtsov05}. The starting
point is action can be split into an unperturbed action $S_0$ and an interaction part W. By Taylor expansion of
partition function in powers of the interaction U, we can reexpress the partition
function
\begin{equation}
Z=\sum_{k,\sigma}Z_0 \frac{(-U)^k}{k!}\int dr_{1\sigma} \dots dr_{k\sigma}
D^{r^{\prime}_{1\sigma} \dots r^{\prime}_{k\sigma}}_{r_{1\sigma} \dots r_{k\sigma}} 
\end{equation}
with the correlation function.
\begin{equation}
D^{r^{\prime}_{1\sigma} \dots r^{\prime}_{k\sigma}}_{r_{1\sigma} \dots
  r_{k\sigma}} = \langle T(c^{\dagger}_{r^{{\prime}_{1\sigma}}}c_{r_{1\sigma}} \dots
  c^{\dagger}_{r^{{\prime}_{k\sigma}}}c_{r_{k\sigma}}) \rangle
\end{equation}      
where $Z_0=Tr(Te^{-S_0})$ is the partition function for unperturbed system,
integration over $dr$ implies the integral over $\tau$ and sum over all
lattices states and $T$ is the time-ordering operator. By Wick's theorem the weight function
$D^{r^{\prime}_{1\sigma} \dots r^{\prime}_{k\sigma}}_{r_{1\sigma} \dots
  r_{k\sigma}}$ is determined by
\begin{equation}
D^{r^{\prime}_{1\sigma} \dots r^{\prime}_{k\sigma}}_{r_{1\sigma} \dots r_{k\sigma}} = \det|g_0(r_i - r_j)| : i,j=1,\dots, k
\end{equation}
where $g_0(r_i - r_j)$ is the bare Green's function. The Green's function is defined by
\begin{equation}
G(r,r^{\prime})=\frac{\langle
  Tc^{\dagger}_{r^{\prime}}c_{r}c^{\dagger}_{r^{{\prime}_{1\sigma}}}c_{r_{1\sigma}}
  \dots
  c^{\dagger}_{r^{{\prime}_{k\sigma}}}c_{r_{k\sigma}} \rangle}{\langle c^{\dagger}_{r^{{\prime}_{1\sigma}}}c_{r_{1\sigma}}
  \dots c^{\dagger}_{r^{{\prime}_{k\sigma}}}c_{r_{k\sigma}}\rangle}
\end{equation}
and by the fast-update formula \cite{Rubtsov05, Haule07} and the Fourier transformation we can rewrite
the Green's function in the Matsubara frequencies space.
\begin{equation}
G(\omega)=g_0(\omega)-g_0(\omega)[\frac{1}{\beta} \sum_{i,j}M_{i,j}e^{i\omega
  (\tau_i - \tau_j)} ]g_0(\omega)
\end{equation}
where $M=D^{-1}$ and $g_0(\omega)$ is the bare Green's function. A
two-particle Green's function related to a vertex function for DF method can
be calculated by Wick's theorem. With this method it is possible to perform
calculations in low temperature regions which cannot be accessed easily with
determinant QMC method without Trotter error. For example, the matrix size of the
CT QMC method is scaled by k $\thicksim 0.5N_cU\beta$ which is comparable to
determinant QMC \cite{Hirsch86} method scaled by k $\thicksim
5N_cU\beta$. Moreover, even if the recently developed strong-coupling CT QMC
method, \cite{Werner06, Haule07} which is
based on a diagrammatic expansion in the impurity-bath hybridization, has nice
advantages such as removing the fermionic sign problem and calculating lower
temperature regions, its computational effort in large cluster is increased exponentially by
the number of sites $N_c$. However, our CT QMC method can overcome the
problem because the computational burden only increases linealy with the
number of sites.

\section{RESULTS}\label{Results}
\subsection{Comparison of the spectral functions for the SCA and C-T QMC methods}
First we compare the one-particle spectral functions obtained from the SCA and CT QMC
methods with Pade approximation for analytical continuation. Since the process, in which $G(i\omega)$ calculated by the CT QMC
method changes into $G(\omega)$ with Pade approximation, introduces large
error, it is useful check to compare
QMC results to SCA results which are calculated in real frequency space. Moreover,
because the systems with geometrical frustration have large $U_c$, the SCA
method is suitable for this model. The spectral function is given by
\begin{equation}
A_{\sigma}(K,\omega) = -\frac{1}{\pi}{\rm Im} G_{\sigma} (K,\omega)
\end{equation}
We compare the Green's function on the Fermi surface K=$(\pi, \pi /
\sqrt{3})$. The results are shown in Fig. 2(a)-(b). At U=6 in Fig. 2(a) the
difference of both results is that the peak of quasiparticle obtained from QMC
method lies around the Fermi level($\omega$=0) due to the geometrical
frustration. On the other hand, the peak obtained from SCA method
deviates from the Fermi level because the SCA underestimates the $\omega$=0
peak. At U=9 in Fig. 2(b) the agreement of
both results is more reasonable and a (pseudo)gap structure is represented.     
\begin{figure}[htp]
\includegraphics[width=250pt]{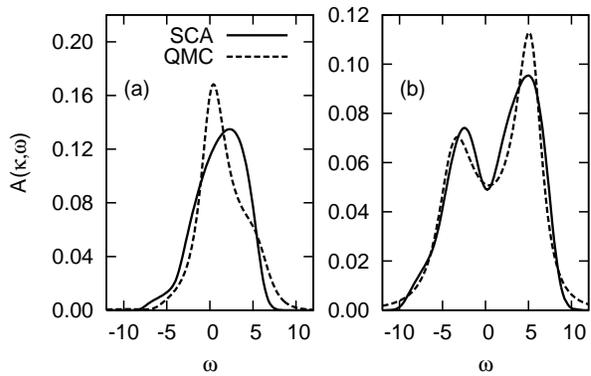}
\caption{One-particle spectral function A(K, $\omega$) corresponding to the
  K=($\pi$,$\pi / \sqrt{3}$) for $\beta$ = 1.6667, (a)U=6 and (b)U=9 by means of the
  SCA and CT QMC with Pade approximation.}
\label{fig:Spectral}
\end{figure}

\subsection{The metal-insulator transition with a first-order transition}
 Here we present our results on the MIT due to geometrical frustration effect
 obtained with the CT QMC method. In previous study of unfrustrated square lattice using DCA method,
it was shown that short-range AF correlations destroy the Fermi liquid
quasiparticle peak at finite temperature \cite{Moukouri01}. According to
Ref. 25, the authors increased the system size gradually in the weak-coupling regime 
on unfrustrated square lattice and measured the total
DOS. Eventually, even if the quasiparticle peak is clearly visible at
$N_c$=1 in DMFT method, there is a small gap at $N_c$=16 which completely disappears at
$N_c$=64. In this system we did not find a band insulator transition. However, on the triangular
lattice the frustration is enough to destroy the AF
correlation. In Fig. 3(a)-(b), we can see the MIT by comparing the total DOS for
U=6, U=10 and $\beta$=4 using the DCA method with $N_c$=4 and
$N_c$=16. Unlike the results for unfrustrated square lattice, the quasiparticle
peak around the Fermi level is clearly seen with increasing $N_c$ at U=6 in
Fig. 3(a). At U=10 in Fig. 3(b) we can see the Mott insulator in both $N_c$=4
and $N_c$=16. This is evidence of a MIT on the triangular lattice.  
\begin{figure}[htp]
\includegraphics[width=250pt]{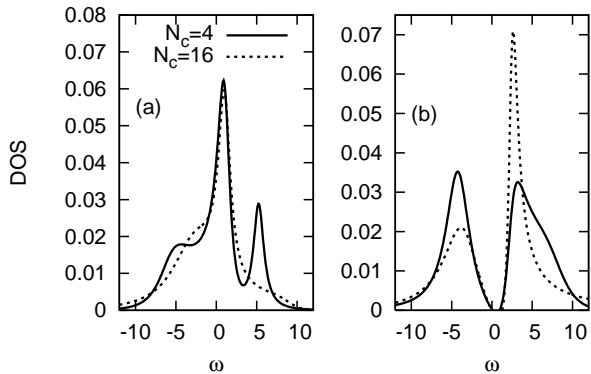}
\caption{Total DOS with $N_c$=4 and $N_c$=16 for $\beta$=4, (a)U=6 and (b)U=10
via CT QMC with Pade approximation.}
\label{fig:Density}
\end{figure}
In the low temperature regime we are also interested in finding whether there is a
first-order transition or a continuous transition and how the geometrical
frustration effects the system. We expect our system to have a first-order
transition because of the recent two cellular DMFT (CDMFT) results \cite{Park08,Ohashi08} which show a first-order
transition on the square lattice with $N_c$=4 and on the triangular lattice with anisotropic
hoping at low temperatures. In order to find evidence of a first-order
transition we measure the double occupancy at several temperatures. Our result is
shown in Fig 4. The system displays a crossover between metal and insulator
at T=0.2. At T=0.1 we can see hysteresis associated with a
first-order transition and at lower temperature hysteresis is more
clear. 
\begin{figure}[htb]
\includegraphics[width=200pt]{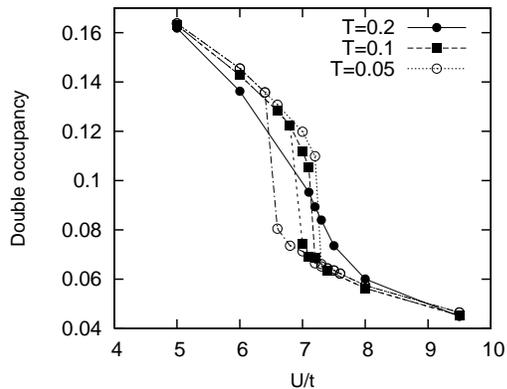}
\caption{Double occupancy as a function of U/t at several
  temperature. $U_c$=7.2 for T=0.2, $U_c$=6.9 for T=0.1 and $U_c$=6.7 for T=0.05.}
\label{fig:double}
\end{figure}
In order to understand the system more clearly we calculate the nearest-neighbor spin
correlation function $\langle S^z_i S^z_{i+1} \rangle$ which is shown in
Fig. 5. At $U_c$ jumps of the spin correlation function indicate the MIT arising from
competetion between the quasiparticle formation and the frustrated spin
correlation. Specifically, the spin correlation is enhanced weakly at $U_c$=7.2 for T=0.2 while
it is increased rapidly at $U_c$ in T $<$ 0.1. Here is $U_c$=6.9 for T=0.1 and
$U_c$=6.7 for T=0.05. This means that the entropy at T=0.2 and in T $<$
0.1 is released by geometrical frustration and spin correlation, respectively
as temperature decreases and the entropy at insulator state in T $<$ 0.1 has small value which is
triggered a first-order transition because of a formation of AF state.   
\begin{figure}[htb]
\includegraphics[width=200pt]{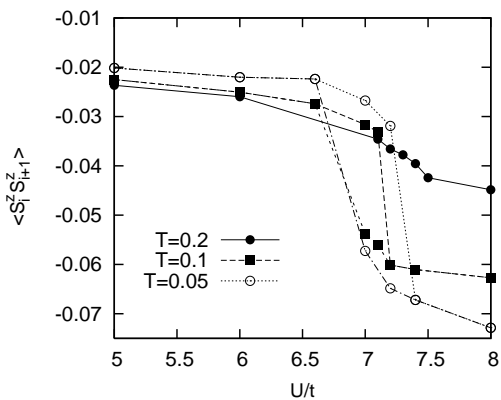}
\caption{The nearest neighbor spin correlation function as a function of U/t
  at several temperature. $U_c$=7.2 for T=0.2, $U_c$=6.9 for T=0.1 and
  $U_c$=6.7 for T=0.05}
\label{fig:spin}
\end{figure}
Moreover, we find that the anomalous character in the metallic
state is unlike the results of the nearest-neighbor spin
correlation on the Kagome lattice\cite{Ohashi06}. In the metallic state the spin correlation
is weak. This is the reason that geometrical
frustration is more dominant than AF spin correlation at lower temperature in the
metallic state because the
frustration on the triangular lattice is stronger than that on the Kagome lattice. However, in the
insulating state AF spin correlation is enhanced stronger than the frustration effect at lower
temperature so the spin correlation is strong with decreasing temperature.     
\subsection{Comparison of Green's functions among the DCA, DF and DMFT methods}
 In this part we using the DMFT, DF and DCA methods with $N_c$=4 and $N_c$=16 to study
 the non-local correlation effects and compare the on-site and
 nearest-neighbor Green's functions in the Matsubara space.
 \begin{figure}[htp]
 \includegraphics[width=250pt]{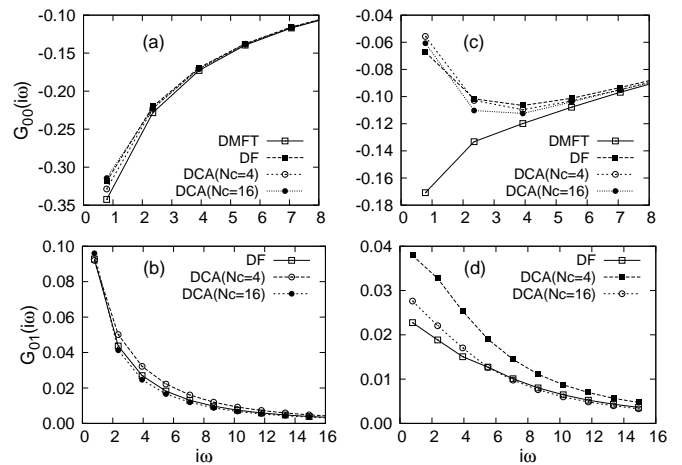}
 \caption{The imaginary part of on-site Green's function for $\beta$=4, (a)
   U=6 and (b) U=10. The real part of nearest-neighbor Green's function for
   $\beta$=4, (c) U=6 and (d) U=10.}
\label{fig:Greenfunction}
\end{figure}
 
 In Fig. 6(a)-(d), we present the Green's functions obtained from DMFT, DF and DCA method with
 $N_c$=4 and $N_c$=16 for $\beta$=4, U=6 and U=10. The on-site Green's
 function of DMFT method in Fig. 6(a) is similiar to the results of DCA
 and DF method and all of these indicate the metallic states. A remarkable point
 is that in Fig. 6(a) and (b), both the on-site and nearest-neighbor Green's functions
 obtained from DF method are closer to those of the DCA method with $N_c$=16
 than $N_c$=4. In Fig. 6(c) at U=10 the on-site Green's function calculated by the
 DMFT still shows the metallic state which overestimates the value of $U_c$ because
 of a lack of non-local correlation. However, the DCA and DF methods can capture
 the insulating state and the agreement of the on-site Green's function
 calculated by the DF and DCA methods with $N_c$=16 is quite reasonable. In Fig. 6(d),
 the nearest-neighbor Green's functions obtained from
 DF method are still closer to those of $N_c$=16 than $N_c$=4. This suggests
 that despite the fact that the DF method is a perturbative method, it would describe physics quite well than the DCA method
 with small cluster size. We expect that considering high-order diagrams will improve the results of the DF method. 
\subsection{The spin susceptibility using the DF method}
In order to explore a magnetic instability we measure the spin
susceptibility using the DF method. The reason why we employ DF method for
the spin susceptibility is that the cluster-extension method of the DMFT takes 
a large amount of time in order to obtain the two-particle properties. On the other
hand, because the DF method includes the vertex renormalization through the
Bethe-Salpeter equation, the computational burden is not serious and the results are
relatively good compared to those of QMC method \cite{Gang08}. Fig. 7(a) shows
$\chi (q)$ for U/t=10.0 and $\beta t=2.5$ where the system is in the
insulating state. The $\chi (q)$ has a maximum peak at $q=(2\pi / 3, 2\pi /
3)$. The spin susceptibility $\chi (q)$ at $q=(2\pi / 3, 2\pi / 3)$ and
q=(0,0) as a function of temperature is exhibited in Fig. 7(b). As temperature
decreases, $\chi (q)$ at $q=(2\pi / 3, 2\pi / 3)$ shows strong enhancement of the AF correlations.

\begin{figure}
\includegraphics[width=250pt]{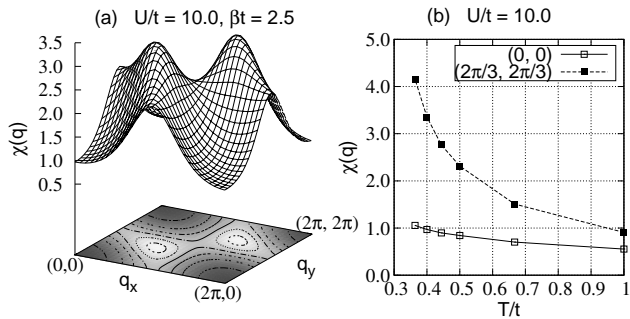}
\caption{(a) The spin susceptibility $\chi (q)$ in the insulating state for
  U/t=10.0 and $\beta t=2.5$. (b) The spin susceptibility as a function of
  temperature at q=(0,0) and $q=(2\pi / 3, 2\pi / 3)$.}
\label{fig:chi}
\end{figure}           

\section{CONCLUSIONS}\label{Summarize}
In summary, we have investigated the Hubbard model on the triangular lattice
using the DCA and DF method. Using the DCA method we compared the spectral functions
obtained from SCA and CT QMC methods. We found a good agreement of both methods and the quasiparticle
peak and gap structure are presented in the weak and the strong coupling
regions, respectively. We found a MIT with a first-order transition at low temperatures
because of the effect of geometrical frustration. Moreover, we employed the DF method which considers the long
range as well as short range correlations
and compared the Green's functions of the DF method to those of the DMFT and DCA method with $N_c$=4 and
$N_c$=16. We found that the DF method does not only overcome the overestimation of
$U_c$ in DMFT method but also that its results are closer the case of $N_c$=16
than $N_c$=4. Finally, we calculated the spin susceptibility $\chi (q)$ via DF
method. We found that the $\chi (q)$ at $q=(2\pi / 3, 2\pi / 3)$ grows rapidly
as temperature decreases.

We would like to thank A. Lichtenstein and E. Gorelov for their assistance in implementing
the CT QMC code.

\bibliography{paper}

\end{document}